\documentclass
[twocolumn,showpacs,preprintnumbers,amsmath,amssymb]{revtex4}

\usepackage{graphicx}
\usepackage{dcolumn}
\usepackage{bm}
\usepackage{rotating}
\usepackage{color}
\usepackage{soul}

\begin{document}
\title{Prospects to attain room temperature superconductivity}
\date{\today}
\author{X. H. Zheng$^1$}
\email{xhz@qub.ac.uk} 
\author{J. X. Zheng$^2$} 
\author{D. G. Walmsley$^1$}
\affiliation{$^1$Department of Physics, Queen's University of Belfast, BT7 1NN, N.  Ireland}
\affiliation{$^2$Department of Electrical and Electronic Engineering, Imperial College London, SW7 2AZ, England}

\begin{abstract}
With a generic model for the electron-phonon spectral density, two simple expressions are derived to estimate the transition temperature and gap-to-temperature ratio in conventional superconductors.  They entail that on average the numerical value of the phonon exchange factor, $\lambda$, is limited to 2.67, so that room temperature superconductivity may be attained only with a Debye temperature of about 1800 K or higher, in materials that may or may not involve hydrogen.  They also show that a Be-Pb alloy may become a superconductor at $\sim$44 K.
\end{abstract}
\pacs{Analytic formulae, Transition temperature, Superconductivity} 
\maketitle

\section{introduction}\label{sec:1}
In 1968 Ashcroft suggested that metallic hydrogen can be a high-temperature superconductor \cite{Ashcroft}.  In 2004 Ashcroft suggested that hydrogen dominant metallic alloys in diamond anvil cells can become high-temperature superconductors at pressures considerably lower than may be necessary for metallic hydrogen \cite{Ashcroft2}.  In 2015 Eremets and co-workers indeed observed that sulfur hydride in a diamond anvil cell becomes a superconductor at 203 kelvin at about 150 GPa \cite{Drozdov, Capitani}.  Here we attempt to clarify whether the prospect of room temperature superconductivity is achievable in principle.
We also demonstrate that, via virtual crystal approximation, a Be-Pb alloy may become a superconductor at $\sim$44~K.

\section{theory}\label{sec:2}
We wish to find a general relation between the superconducting transition temperature, $T_c$, and other properties of the material, including the phonon exchange factor, $\lambda$, and Debye temperature, $T_D$, from numerical solutions to the Eliashberg equations.  We apply the following generic model to evaluate the equations:
\begin{eqnarray}\label{eq:1}
\alpha^2F(\nu) = \lambda\nu^2/(k_BT_D)^2,\;\;\;\;\;\;\mbox{if}\;\;\nu\leq k_BT,
\end{eqnarray}
otherwise $\alpha^2F(\nu) = 0$, where $k_B$ is the Boltzmann constant, and $\nu$ the phonon frequency in joule or eV, see the Appendix for a justification.  It is also consistent with the definition of $\lambda$ \cite{McMillan} because, when we integrate $2\alpha^2F(\nu)/\nu$ over $\nu$, we simply recover the value of $\lambda$.  In Eq.~(\ref{eq:1}) the values of $\lambda$ and $T_D$ embody all phonon properties, leaving no ambiguities to our discussion within its premise. 

In Eq.~(\ref{eq:1}) $T_D$, if not available, can be replaced by the characteristic temperature, $\Theta$, in the Bloch-Gr\"uneisen formula for electrical resistivity, which is usually close to $T_D$, where $\Theta = 1223$~K in H$_3$S, found from Eq.~(\ref{eq:A4}) and FIG.~S3 in \cite{Capitani} via numerical fitting.  We also let the Coulomb pseudopotential $\mu^*$ = 0.09, 0.13 and 0.17, a usual range encountered experimentally \cite{Allen, Mitrovic}.  We will let $T = T_c$ once $\Delta(T) = 0.1\Delta_0$ instead of $\Delta(T) = 0$ {\cite{McMillan}}. This amounts to a new and just as accurate definition of $T_c$, because $\Delta(T)$ vanishes quickly as soon as it reaches $0.1\Delta_0$ with increasing $T$, a fact well known from the exemplary $\Delta(T)$ curve in \cite{BCS}.

\section{Results}\label{sec:3}
\begin{figure}[h]
\includegraphics[width=8cm]{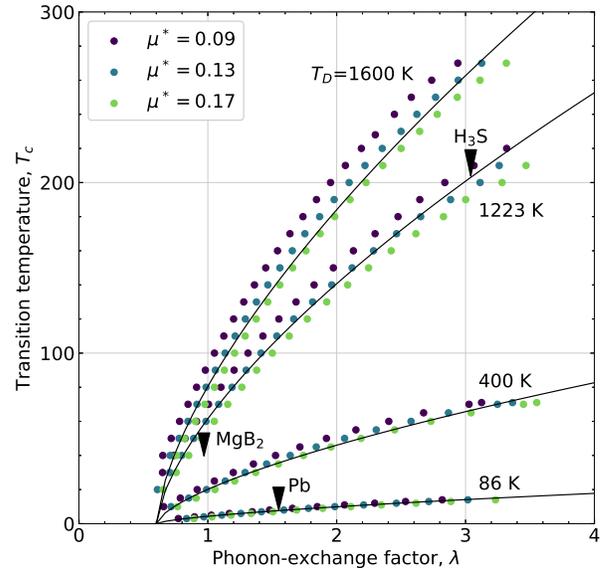}
\caption{Transition temperatures (dots) from the generic model in Eq.~(\ref{eq:1}) against $\lambda$, $\mu^* = 0.09$, 0.13 and 0.17 (colour coded on line), $T_D = 86$, 400, 1223 and 1600 K.  Experimental data for Pb, MgB$_2$ and H$_3$S are marked by downward black arrow heads.  The curves are from Eq.~(\ref{eq:2}).}
\label{fig:1}
\end{figure}

We are now adequately equipped to solve the Eliashberg equations.  Usually the procedure starts with a destination value of $\Delta_0$, giving $T_c$ after the procedure terminates {\cite{McMillan+Rowell}}; we instead start with a destination value of $T_c$.  With given values of $T_D$ and $\mu^*$, we find $\lambda$ via an optimization program of two dimensional search.  In one dimension we search $T$ to let the Eliashberg equations produce $\Delta(T) = 0.1\Delta_0$.  In the other dimension we search $\lambda$ until $|T - T_c|\sim10^{-7}$~K.  Meanwhile $\Delta_0$ keeps evolving until our procedure terminates.

\begin{figure}
\resizebox{8cm}{!}{\includegraphics{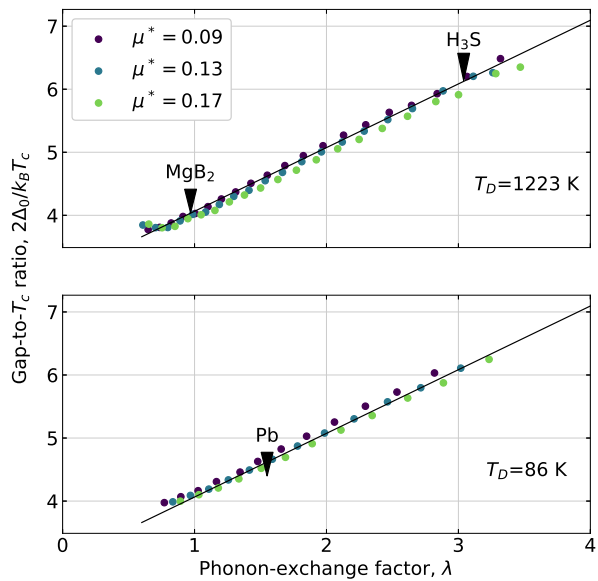}}
\caption{Gap-to-$T_c$ ratio, $2\Delta_0/k_BT_c$ (dots), from the generic model in Eq.~(\ref{eq:1}), against $\lambda$, $\mu^* = 0.09$, 0.13 and 0.17 (colour coded on line), $T_D = 86$ (lower) and 1223 K (upper).  Data for Pb, MgB$_2$ and H$_3$S are marked by downward black arrow heads.  The lines are from Eq.~(\ref{eq:3}).}
\label{fig:2}
\end{figure}

Our results, shown as dots in FIG.~\ref{fig:1}, align themselves into narrow bands, largely regardless of the values of $\mu^*$.  In particular, when $T_D = 86$ and 400 K, the dots appear to move along the same curve, in response to the varying values of $\mu^*$ --- we therefore have an expression for $T_c$ in terms of only $\lambda$ and $T_D$, without involving $\mu^*$ but still retaining reasonable accuracy.  We find the following power law relationship from numerical fitting
\begin{eqnarray}\label{eq:2}
T_c = 0.23\,T_D\Big(0.25\,\lambda - 0.15\Big)^{0.66}
\end{eqnarray}
which is applicable when $\lambda\ge0.6$.  We also find the following linear relationship
\begin{eqnarray}\label{eq:3}
\frac{2\Delta_0}{k_BT_c} = 4.04\,\Big(0.25\,\lambda - 0.15\Big) + 3.66
\end{eqnarray}
which is likewise applicable when $\lambda\ge0.6$.  Outcomes of Eqs.~(\ref{eq:2}) and (\ref{eq:3}) are shown as curves and lines in FIGs.~\ref{fig:1} and \ref{fig:2} respectively.

\begin{figure}[h]
\begin{minipage}{8cm}
TABLE I: SUPERCONDUCTOR PARAMETERS
\vspace{2mm}
\begin{ruledtabular}
\begin{tabular}{ccccc}
Matter&$T_D\footnotemark[1]$&$T_c$\footnotemark[1]&$\lambda$\footnotemark[2]&$2\Delta_0/k_BT_c$\footnotemark[2]\\ \hline
V&380&5.38&0.66 (0.81)&3.72 (3.40)\\
Sn&200&3.72&0.69 (0.72)&3.75 (3.50)\\
Ta&240&4.48&0.69 (0.69)&3.75 (3.60)\\
In&108&3.40&0.80 (0.81)&3.86 (3.60)\\
Nb&275&9.50&0.83 (0.97)&3.89 (3.80)\\
MgB$_2$\footnotemark[3]&815&39.0&0.97\,(0.92)&4.03\,(4.02)\\
Hg&71.9&4.16&1.09 (1.62)&4.16 (4.60)\\
Pb&105&7.19&1.24 (1.55)&4.30 (4.38)\\
Li\footnotemark[3]&147&14.9&1.76 \,(NA)&4.83 \,(NA)\\
H$_3$S\footnotemark[3]&1223&203&3.04 \,(NA) & 6.12 \,(NA)
\end{tabular}
\end{ruledtabular}
\end{minipage}\hfill
\footnotetext[1]{From \cite{Kittel} unless stated otherwise.}
\footnotetext[2]{Calculated or directly from \cite{Mitrovic, Masui, Dolgov, Kittel} (bracketed).}
\footnotetext[3]{Calculated, $T_c$ from \cite{Drozdov, Masui, Shimizu}.}
\end{figure}

\begin{figure}
\resizebox{8cm}{!}{\includegraphics{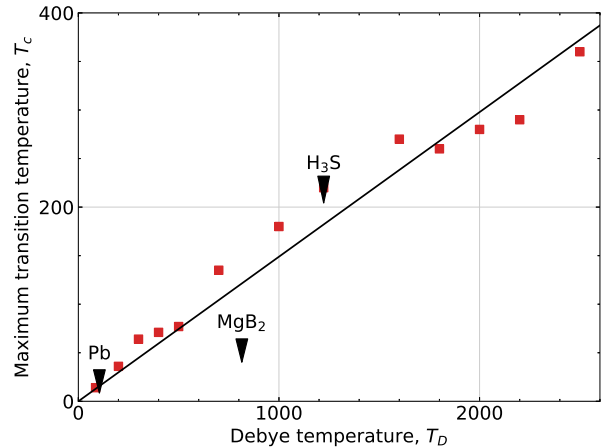}}
\caption{Observed maximum values of $T_c$ (in kelvin, filled squares) and $T_c$ in Pb, MgB$_2$ and H$_3$S (downward arrow heads) against $T_D$.  The line is from Eq.~(\ref{eq:2}) with $\lambda = 2.67$.}
\label{fig:3}
\end{figure}

We always encounter an upper limit of $\lambda$ in our numerical procedure. For example, at $T_D = 400$ K, our numerical procedure fails to converge once $\lambda > 2.57$, or $T_c > 77$~K, which we will refer to as the observed maximum value of the transition temperature (possible reasons discussed in Section~\ref{sec:5}). In FIG.~\ref{fig:3} we plot the empirical maximum values of $T_c$ (filled squares) found at $86\leq T_D\leq2500$ K.  We also plot a line for $T_c$ from Eq.~(\ref{eq:2}), with $\lambda = 2.67$ to ensure a minimum r.m.s.~difference between the output of the formula and values of the observed maximum $T_c$.

In TABLE~1 the values of $T_D$ in dense Li and H$_3$S are replaced by the Bloch-Gr\"uneisen characteristic temperatures \cite{Ziman} we extracted from the electrical resistance data in \cite{Drozdov, Shimizu}.  MgB$_2$ is a two band superconductor, with two values of $\Delta_0$, and we evaluate $2\Delta_0/k_BT_c$ with the larger $\Delta_0$ \cite{Masui}.   The experimental $\lambda$ in MgB$_2$ is an average between its values in the $\sigma$ and $\pi$ bands \cite{Dolgov}.  The theoretical $\lambda$ is calculated from Eq.~(\ref{eq:2}) with experimental $T_D$ and $T_c$.  The theoretical gap-to-$T_c$ ratios are from Eq.~(\ref{eq:3}) with theoretical $\lambda$.

To the best of our knowledge experimental values of $\lambda$ and $2\Delta_0/k_BT_c$ in dense Li and H$_3$S are not yet available.  According to Errea and co-workers $\lambda = 2.64$ or 1.84 for harmonic or anharmonic phonons respectively in H$_3$S in theory, with $\Delta_0\approx43$ meV and $T_c = 190$ K ($\mu^* = 0.16$, anharmonic phonons) giving $2\Delta_0/k_BT_c\approx5.25$ \cite{Errea}, compared with $\lambda$ between 2.07 and 2.19 in (H$_2$S)$_2$H$_2$ in theory \cite{Duan}, largely consistent with our results considering the many uncertainties in theoretical evaluations.

\section{McMillan formula}\label{sec:4}
\begin{figure}[h]
\includegraphics[width=8cm]{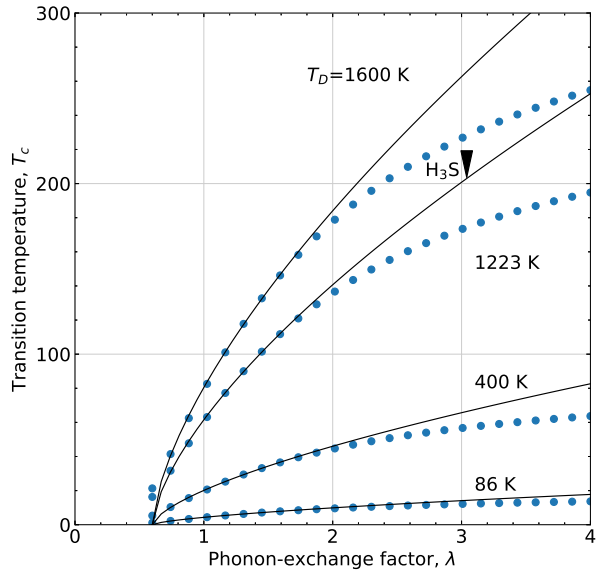}
\caption{Comparison between transition temperatures from Eq.~(\ref{eq:2}) (curves) and Eq.~(\ref{eq:4}) (dots, $\mu^* = 0.13$).}
\label{fig:4}
\end{figure}

In 1968 McMillan \cite{McMillan} solved the Eliashberg equations via iteration with a number of simplifications.  The equations are linearised at $T = T_c$, the solutions are assumed to have just two values, $\Delta_0$ and $\Delta_\infty$, defined immediately beneath the surface and deeply inside of the Fermi sphere, respectively, and $\alpha^2F(\nu) = \alpha^2\times F(\nu)$, $\alpha$ being a constant and $F(\nu)$ the phonon density of states (from Nb neutron scattering experiment for any bcc lattice, assumed to vanish below 8.6 meV).  During the iteration $T_c$ and $\mu^*$ are kept constant, $\alpha$ adjusted continuously to keep $\Delta_0$ constant.  The formula
\begin{eqnarray}\label{eq:4}
T_c = \frac{T_D}{1.45}\exp\left[-\frac{1.04(1 + \lambda)}{\lambda - \mu^*(1 + 0.62\lambda)}\right]
\end{eqnarray}
results from numerical fitting, and has proven to be highly successful in guiding both theoreticians and experimentalists.  For clarity we choose $\mu^* = 0.13$ and plot the outcome of Eq.~(\ref{eq:4}) in FIG.~\ref{fig:4} as dots, which matches the outcome of Eq.~(\ref{eq:2}) closely if $\lambda<2$, but otherwise underestimates $T_c$, rather significantly in the parameter region of H$_3$S. 

In 1975 Allen and Dynes \cite{Allen} published a slightly modified $T_c$ formula where $T_D/1.45$ in Eq.~(\ref{eq:4}) is replaced by $\omega_{\mbox{\scriptsize ln}}/1.2$, $\omega_{\mbox{\scriptsize ln}} = (2/3)T_D$ with our generic model in Eq.~(\ref{eq:1}) in place, that is $T_c$ from Eq.~(\ref{eq:4}) will be suppressed by $\sim$20\%.  In 1984 Mitrovi\'c, Zarate and Carbotte \cite{Mitrovic} found an approximate formula for $2\Delta_0/k_BT_c$ in terms of $T_c/\omega_{\mbox{\scriptsize ln}}$, not directly comparable with Eq.~(\ref{eq:3}).

\section{beryllium-lead alloy}\label{sec:6}

\begin{figure}[h]
\resizebox{8cm}{!}{\includegraphics{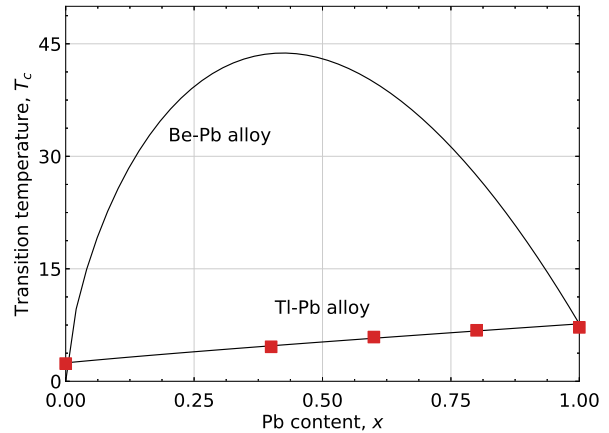}}
\caption{Transition temperatures in Tl-Pb alloys.  The dots are experimental data and the curves are theoretical calculations from Eq.~(\ref{eq:2}) with the virtual crystal approximation.}
\label{fig:5}
\end{figure}

Eqs.~(\ref{eq:2}) and (\ref{eq:3}) enable us to predict superconductivity quantitatively, not only in simple metals but also in alloys.  For example, in Pb we have $T_D\simeq\Theta_1 = 86$ K (Bloch-Gr\"uneisen characteristic temperature) and $\lambda = \lambda_1 = 1.55$, giving $T_c = 7.66$ K via Eq.~(\ref{eq:2}), compared with an empirical value of 7.19 K \cite{Mitrovic}.  In Tl we have $T_D = \Theta_2 = 78.5$ K~\cite{Kittel} and $\lambda = \lambda_2 = 0.795$, giving $T_c = 2.46$ K, compared with the empirical 2.36 K \cite{Mitrovic}.  Letting $T_D = x\Theta_1 + (1 - x)\Theta_2$ and $\lambda = x\lambda_1 + (1 - x)\lambda_2$, where $x$ is the content of Pb in a Tl-Pb alloy in accordance with the well-established virtual crystal approximation~\cite{Nordheim}, we find $T_c$ in the alloys from Eq.~(\ref{eq:2}) with r.m.s.~deviation = 4.2\% against experimental data, as is shown in FIG.~\ref{fig:5} and TABLE~2.

\begin{figure}[h]
\begin{minipage}{8cm}
TABLE II: THALLIUM-LEAD ALLOY PROPERTIES\footnotemark[1]
\vspace{2mm}
\begin{ruledtabular}
\begin{tabular}{ccccc}
Pb&Tl&$\Theta$&$\lambda$&$T_c$\footnotemark[2]\\ \hline
0.0&1.0&78.5&0.795&2.46 (2.36)\\
0.4&0.6&81.5&1.097&4.73 (4.60)\\
0.6&0.4&83.0&1.248&5.74 (5.90)\\
0.8&0.2&84.5&1.399&6.71 (6.80)\\
1.0&1.0&86.0&1.550&7.66 (7.19)
\end{tabular}
\end{ruledtabular}
\end{minipage}\hfill
\footnotetext[1]{From virtual crystal approximation.}
\footnotetext[2]{Experimental values from \cite{Mitrovic} (bracketed).}
\end{figure}

Replacing Tl with Be allows us to now make a prediction for $T_c$ in Be-Pb alloys, by applying the same virtual crystal approximation. With $\Theta_3 = 1440$~K (Debye temperature) and $\lambda_3 = 0.6$ to let $T_c\sim0$ (= 0.026 K in reality) for Be~\cite{Kittel}, we let $T_D = x\Theta_1 + (1 - x)\Theta_3$ and $\lambda = x\lambda_1 + (1 - x)\lambda_3$. This leads through Eq.~(\ref{eq:2}) to give the upper curve in FIG.~\ref{fig:5}, where the value of $T_c$ reaches a peak of 43.8~K. 

\section{discussion}\label{sec:5}
Eqs.~(\ref{eq:2}) and (\ref{eq:3}) entail a few interesting physical consequences.  First, there must be a upper limit of $\lambda$.  If $\lambda\gg1$ then we have $\Delta_0\sim0.51\lambda k_BT_c$ from Eq.~(\ref{eq:3}), and this leads through Eq.~(\ref{eq:2}) to $\Delta_0\sim0.47\lambda^{1.66}k_BT_D$. Consequently we have $\Delta_0/\lambda k_BT_D\sim0.47\lambda^{0.66}\rightarrow\infty$ when $\lambda\rightarrow\infty$. This is impossible, because $\Delta_0$ is the edge of superconducting energy gap function, whereas $\lambda k_BT_D$ measures the strength of the electron-phonon interactions, the underlying reasons for superconductivity to arise, and the two must be compatible.  This explains why in FIGs.~\ref{fig:1}, \ref{fig:2} and \ref{fig:3} the numerical values of $\lambda$ are limited to 2.67 on average. 

It is remarkable that McMillan also found a value of $\lambda$ in association with maximum $T_c$ \cite{McMillan}.  He argued that in Eq.~(\ref{eq:4}) $T_D\sim\langle\omega\rangle$ and $\lambda\sim1/\langle\omega^2\rangle$, where $\omega$ stands for phonon frequency, so that $T_c\rightarrow0$ when $\langle\omega\rangle$ is either too large or too small.  Searching for maximum $T_c$ leads to $\lambda = 2$, not far from our value 2.67.  He estimated $T_c$ may reach 9.2 K for lead-based alloys and 40 K for V$_3$Si \cite{McMillan}. It appears that $T_c$ in the BCS theory is indeed curbed by an intrinsic limit, which cannot be lifted by strong electron-phonon interactions, reminiscent of the speed limit imposed by the Lorentz transformation.

Second, it is immediately apparent from Eq.~(\ref{eq:2}) that, with $\lambda\sim2.67$, the only realistic try for us to achieve high $T_c$ is in materials with high $T_D$.  Indeed we know from TABLE I that in MgB$_2$ we have $T_c = 39$ K with $\lambda = 0.97$ and $T_D = 815$ K, compared with $T_c = 7.19$ K, $\lambda = 1.24$ but $T_D = 105$ K in Pb. If somehow we were able to let $\lambda = 2.67$ in both cases, then in MgB$_2$ we would have $T_c = 121$ K from Eq,~(\ref{eq:2}), compared with $T_c = 16$ K in Pb.  Indeed, it is clear from FIG.~\ref{fig:1} that, ascending along the $T_c$ curves with $T_D = 86$ or 400 K, there is hardly any hope for us to reach $T_c\sim300$ K.

Third, we might be able to push $T_c$ into the range of room temperatures, but just barely.  We know from TABLE I that we may have $\lambda = 3.04$, and this leads through Eq.~(\ref{eq:2}) to $T_c = 300$ K with $T_D\sim1800$ K.  From direct numerical calculation we find $T_c = 255$ K with $\lambda = 2.39$ and $T_D = 1800$ K.  Whether or not we can achieve higher $T_c$ with some $\alpha^2F(\nu)$ other than the generic model in Eq.~(\ref{eq:1}) remains an open question, but we suspect the potential is rather limited, unless $T_D\sim2500$ K or higher were an option at our disposal. 

Finally, the close fit between the dots and the lower curve in FIG.~5 tells us that $T_c$, from Eq.~(\ref{eq:2}), in virtual crystal approximation, is trustworthy in the case of the Tl-Pb alloys.  If the higher curve in FIG.~\ref{fig:5} is just as accurate, we may achieve $T_c\sim44$ K in Be-Pb alloys, due to Fermi surface enlargement, on account of the large valence in Pb (= 4). The Be-Bi alloy may also be worth considering because Bi, though not a superconductor in bulk form, was proven to be highly effective to raise $T_c$ in Pb-Bi alloys \cite{Mitrovic} apparently likely due to its very large valance (= 5). 

\section{conclusions}\label{sec:7}
We conclude with suggestions for future work.  We wish for the experimental values of $\lambda$ and $2\Delta_0/k_BT_c$ in dense lithium and H$_3$S to be made available, to validate or dispute our theoretical predictions in TABLE I.  In the periodic table $T_D$ can reach 1440 and 2230 K in beryllium and carbon respectively \cite{Kittel} formally nearly or more than enough for $T_c\sim300$ K with $\lambda\sim3$. Indeed $T_c = 18$ K in potassium-doped C$_{60}$ \cite{Hebard}.  It would be interesting to see what will happen in beryllium, doped with lead or other metals to become an alloy with an enlarged Fermi surface.  Experimentally this may be achieved via the vacuum-sputtering technique \cite{inam} to make a uniform film of an alloy of two or more metals or, if necessary, a pancake film of multiple layers of different atoms.

\setcounter{equation}{0}
\renewcommand\theequation{A\arabic{equation}}
\section*{APPENDIX}
Here we justify Eq.~(\ref{eq:1}) in the main text. By definition
\begin{equation}\label{eq:A1}
\alpha^2F(\nu) = \frac{1}{N}\sum_{\ell, {\bf q}}\delta(\nu - \hbar\omega_\ell)\delta(\epsilon - \epsilon_F)\left|g_\ell({\bf q})\right|^2
\end{equation}
for a spherical Fermi surface, where $\ell$ identifies phonon polarization, $N$ is the number of atoms per unit volume, $\bf q$ and $\omega_\ell = \omega_\ell({\bf q})$ phonon momentum and frequency, $\bf k$ and $\epsilon = \epsilon({\bf k})$ electron momentum and energy, $\epsilon_F$ Fermi energy, and $g_\ell({\bf q})$ matrix element \cite{Truant}.  In the normal state
\begin{eqnarray}\label{eq:A2}
\alpha_{\mbox{\scriptsize tr}}^2F(\nu) = \frac{1}{N}\sum_{\ell, {\bf q}}\delta(\nu - \hbar\omega_\ell)\delta(\epsilon - \epsilon_F)\left|g_\ell({\bf q})\right|^2\nonumber\\\times\left[\frac{\bf k\cdot(k + q)}{\bf k\cdot k} - 1\right]
\end{eqnarray}
and its model
\begin{equation}\label{eq:A3}
\alpha_{\mbox{\scriptsize tr}}^2F(\nu) = \left\{\begin{array}{cc}\lambda_{\mbox{\scriptsize tr}}(\nu/k_B\Theta)^4, & \nu\le k_B\Theta\\\\0, & \mbox{otherwise}\end{array}\right.
\end{equation}
lead through relevant formulations \cite{Tomlinson} to the Bloch-Gr\"uneisen formula 
\begin{eqnarray}\label{eq:A4}
\frac{\rho(T)}{\rho(\Theta)} = 4.2263\left(\frac{T}{\Theta}\right)^{\!5} \int_0^{\,\Theta/T}\!\!\!\frac{x^5dx}{(e^x - 1)(1 - e^{-x})}
\end{eqnarray}
where $\rho(T)$ and $\rho(\Theta)$ are electrical resistivity measured at $T$ and $\Theta$, respectively, $\Theta$ being the characteristic temperature, highly accurate in most metals over a broad range of temperatures \cite{Ziman}.  If the sound velocity is constant anywhere within the first Brillouin zone in reciprocal space (Debye model) then
\begin{equation}\label{eq:A5}
\frac{\bf k\cdot(k + q)}{\bf k\cdot k} - 1\propto\left(\frac{\nu}{k_B\Theta}\right)^{\!\!2}
\end{equation}
for normal phonons.  Assuming that Eq.~(\ref{eq:A5}) also applies to umklapp phonons, that is assuming the size of the phonon sphere is always proportionally measured by the value of phonon frequency (energy), we find Eq.~(\ref{eq:1}) via Eqs.~(\ref{eq:A1}--\ref{eq:A3}).

\section*{ACKNOWLEDGEMENTS}
The authors wish to thank Professor Nikolay Plakida for useful discussions.

\end{document}